\documentclass[11pt]{article}
\usepackage[pdftex]{graphicx,color} 
\usepackage{jheppub}
\usepackage{amsmath}
\usepackage{amssymb}
\usepackage{comment}
\usepackage{shuffle}
\usepackage{multirow}
\usepackage{mathtools}
\usepackage{dsdshorthand}
\newcommand{\bea}{\begin{equation}\begin{aligned}}
\newcommand{\eea}[1]{\label{#1}\end{aligned}\end{equation}}
\newcommand{\beq}{\begin{equation}}
\newcommand{\eeq}{\end{equation}}

\newcommand   \zb  {\bar{z}}

\usepackage{tikz}
\usetikzlibrary{arrows,calc,shapes,decorations.pathmorphing,decorations.markings}
\tikzset{
>=stealth',
help lines/.style={dashed, thick},
axis/.style={<->},
important line/.style={thick},
connection/.style={thick, dotted},
  cross/.style={
    cross out,
    draw=black, 
    minimum size=5pt, 
    inner sep=0pt,
    outer sep=0pt
  },
->-/.style={decoration={
  markings,
  mark=at position #1 with {\arrow{>}}},postaction={decorate}}
}

\title{High Energy String Scattering in AdS}

\author[a]{Luis F. Alday,}
\author[a,b]{Tobias Hansen}
\author[a]{and Maria Nocchi}

\affiliation[a]{Mathematical Institute, University of Oxford,
Woodstock Road, Oxford, OX2 6GG, UK}

\affiliation[b]{Department of Mathematical Sciences, Durham University,
Stockton Road, Durham, DH1 3LE, UK}

\abstract{ We study the AdS Virasoro-Shapiro amplitude in the limit of fixed-angle high energy scattering. A recent representation as a world-sheet integral allows to compute the amplitude in this regime by saddle point techniques, very much as in flat space. This result is then compared to a classical scattering computation in AdS and agreement is found. As a byproduct of this comparison we show that AdS curvature corrections exponentiate in the high energy limit. 
}

\emailAdd{alday@maths.ox.ac.uk}
\emailAdd{tobias.p.hansen@durham.ac.uk}
\emailAdd{maria.nocchi@maths.ox.ac.uk}

\begin{document}
\maketitle

\section{Introduction and main results}

Perturbative string theory in flat backgrounds is relatively well understood. In principle world-sheet methods can be used to compute perturbative expansions for a variety of scattering amplitudes \cite{Berkovits:2022ivl}. The situation in curved spacetime is very different. Currently we lack the technology to compute string scattering amplitudes in general curved backgrounds, even at tree level. In this paper we will focus on  $AdS_5 \times S^5$. Even in this maximally supersymmetric case, the presence of RR backgrounds makes the standard RNS formulation hard to apply\footnote{In principle it is possible to use the RNS formulation to compute amplitudes in a curvature expansion around flat space. We expect this computation to become very cumbersome very quickly.}, while other formulations, such as the Green-Schwarz or pure spinor, are not yet at the point where they can yield results for scattering amplitudes. Very recently progress has been made for the simplest case: the four-graviton amplitude at tree-level $A_{4}^{AdS}(S,T)$. Tools from $AdS/CFT$, number theory and integrability were combined in \cite{Alday:2022uxp,Alday:2022xwz,Alday:2023jdk,Alday:2023mvu} to compute $A_{4}^{AdS}(S,T)$ as a curvature expansion around flat space. The final result takes the form of a world-sheet integral, analogous to that of the Virasoro-Shapiro amplitude, with extra insertions
 \begin{equation}
 \label{wsrep}
 A_{4}^{AdS}(S,T) \sim \int d^2 z |z|^{-2 S}|1-z|^{-2T} W_0(z,\bar z) \left(1+ \frac{S^2}{R^2}W_3(z,\bar z) + \frac{S^4}{R^4} W_6(z,\bar z) + \cdots\right)\,,
 \end{equation}
 where $W_0(z,\bar z)=\frac{1}{2 \pi U^2 |z|^2 |1-z|^2}$ is a rational function and $W_3(z,\bar z),W_6(z,\bar z)$, explicitly known, are specific single-valued transcendental functions, of weights three and six in the `world-sheet' coordinates $(z, \bar z)$ and we have set $\alpha'=1$. We would like to stress that this is not the result of a direct world-sheet computation, currently unavailable, but rather the rewriting of a result found by completely different methods. A natural question is how we use this result to learn about strings on $AdS_5 \times S^5$ and how to make the connection to a more direct world-sheet computation. In this paper we study a regime in which such a connection, at least classically, can be made. 

String amplitudes in flat space display remarkable properties in the high energy limit $|S|, |T| \gg 1$ with fixed $S/T$, see \cite{Gross:1987kza,Gross:1987ar,Gross:1988ue}. For instance the Virasoro-Shapiro amplitude $ A_{4}^\text{flat}(S,T)$ falls off exponentially in this regime
\begin{equation}
 A_{4}^\text{flat}(S,T) \sim e^{-2 S \log|S| -2 T \log|T| - 2U \log |U|}\,,
\end{equation}
in clear contrast to field theory amplitudes, which either diverge or fall off as a power. This behaviour can be directly understood from the integral representation for $A_{4}^\text{flat}(S,T)$
 \begin{equation}
 A_{4}^\text{flat}(S,T) \sim \int d^2z |z|^{-2 S}|1-z|^{-2T}  W_0(z,\bar z)\,. 
 \end{equation}
Indeed, at high energies $|S|, |T| \gg 1$ the integral can be computed by saddle point approximation around the saddle
\begin{equation}
z = \bar z = \frac{S}{S+T} \equiv z_0\,,
\end{equation}
which leads to 
 \begin{equation}
 A_{4}^\text{flat}(S,T)_\text{HE} \sim W_0( z_0)   e^{-2 S \log|S| -2 T \log|T| - 2U \log |U|}.
 \end{equation}
Let us now make the following interesting remark for the case of $AdS$: given the world-sheet representation (\ref{wsrep}) and because the transcendental functions $W_n( z,\bar z)$ depend on $S,T$ in a polynomial way, the location of the saddle is not modified in a $1/R$ expansion. We can then readily compute $ A_{4}^{AdS}(S,T)$ in the high energy limit
 \begin{equation}
 \label{hedirect}
 A_{4}^{AdS}(S,T)_\text{HE} \sim  e^{-2 S \log|S| -2 T \log|T| - 2U \log |U|}W_0( z_0)  \left(1 + \frac{S^2}{R^2}W_3(z_0) + \frac{S^4}{R^4}W_6(z_0) +\cdots \right)\,,
 \end{equation}
where at each order in $1/R$ we keep the leading large energy contribution. Effectively, we are looking into a regime with large $R,S$ and $S^2/R^2$ finite. 

The high energy regime in flat space can also be understood from the point of view of spacetime, in terms of classical solutions. To do that we consider the path integral representation for the amplitude

 \begin{equation}
 A_{4}^\text{flat}(S,T) \sim \int {\cal D} g {\cal D} X \exp\left( -\frac{1}{4\pi} \int d\zeta_1 d \zeta_2  \sqrt{g} g^{\alpha \beta} \partial_\alpha X^\mu \partial_\beta X_\mu \right) \prod_{i=1}^4 V_i(p_i)\,,
 \end{equation}
where the vertex operators are of the form $V_i(p_i) \sim \int d^2 z_i \sqrt{g} e^{i p_i \cdot X(z_i)}$ with $p_i^2=0$\footnote{In the high energy limit all operators, with a finite number of excitations, are effectively massless.} and the locations $z_i$ are assumed to be real for simplicity. The Mandelstam variables are related to the momenta by $p_1 \cdot p_2 = -2S$, $p_1 \cdot p_3 = -2T$, $p_1 \cdot p_4 = -2U$. At high energies the path integral is dominated by a classical solution 
 \begin{equation}
X_\text{classical}^\mu(\zeta) = -i \sum_k p_k^\mu \log\left|1-\frac{\zeta}{z_k}\right|\,,
 \end{equation}
together with the condition
 \begin{equation}
 \label{zicondition}
z_0 = \frac{S}{S+T} = -\frac{(z_2-z_1)(z_4-z_3)}{(z_3-z_2)(z_4-z_1)}\,,
 \end{equation}
for the location of the punctures. This condition ensures that the induced metric on the world-sheet is conformal. Plugging the classical solution into the path integral reproduces the correct high energy result. In this paper we perform this classical analysis for $AdS$. We study the obvious generalisation to $AdS$ of the path integral
 \begin{equation}
 A_{4,\text{bos}}^{AdS}(S,T) \sim \int {\cal D} g {\cal D} X \exp\left( -\frac{1}{4\pi} \int d\zeta_1 d \zeta_2  \sqrt{g} g^{\alpha \beta} \partial_\alpha X^M \partial_\beta X_M\right) \prod_{i=1}^4 V_i(P_i)\,,
 \end{equation}
where now $X^N$ denote embedding coordinates with $X^M X_M =-(X^0)^2 + X^\mu X_\mu=-R^2$ and the vertex operators are of the form $V_i(P_i) \sim \int d^2 z_i \sqrt{g} e^{i P_i^M X_M(z_i)}$. We expect that at high energies the behaviour of the amplitude is captured by classical solutions, now in $AdS$. Furthermore, we expect that this bosonic model already captures the leading behaviour at high energies, and that other fields, as well as the precise prefactor and details of the vertex operators, will be suppressed by powers of $S$, $T$.\footnote{We will come back to this point in section \ref{sec:universality}.} We give an algorithm to compute the corresponding $AdS$ classical solutions to arbitrary order in a $1/R$ expansion 
\begin{equation}
X_\text{classical}^\mu = X_0^\mu + \frac{1}{R^2} X_1^\mu + \cdots\,.
\end{equation}
$X_0^\mu$ reproduces the solution in flat space. Furthermore, to order $1/R^{2n}$, $X_n^\mu$ is given in terms of single-valued multiple polylogarithms (SVMPLs), whose `letters' are the locations of the punctures $z_i$. More specifically, $X_n^\mu$ has weight $2n+1$. One can also show that the solution is consistent with the Virasoro constraints, provided (\ref{zicondition}) is satisfied. Plugging this classical solution into the action we obtain
 \begin{equation}
 \label{bosonicmodel}
 A_{4,\text{bos}}^{AdS}(S,T)_\text{HE} \sim A_{4}^\text{flat}(S,T)_\text{HE} \times e^{\frac{S^2}{R^2} V_3(z_0) +\frac{S^3}{R^4} V_5(z_0) +\cdots}\,,
 \end{equation}
where $V_3(z_0),V_5(z_0),\cdots$ denote combinations of transcendental functions of weight three, five, and so on. Upon expanding the exponential in powers of $1/R$, powers of the first term will dominate over powers containing also other terms, so that  $\frac{S^3}{R^4} V_5(z_0) +\cdots$ can be safely ignored at leading order in the high energy limit. The same is true about the precise prefactors multiplying the vertex operators. After an appropriate redefinition of the Mandelstam variables $S \to S(1+2 \frac{\alpha}{R^2})$, one can explicitly check that (\ref{bosonicmodel}) correctly reproduces the results in (\ref{hedirect}) in the high energy limit. To be more precise, to each order in $1/R$ this formula captures the leading term at high energy. In particular $W_3(z_0)=V_3(z_0)$ and $W_6(z_0)=\frac{1}{2}V_3(z_0)^2$. We hence arrive at the result
 \begin{equation}
 A_{4}^{AdS}(S,T)_\text{HE} \sim A_{4}^\text{flat}(S,T)_\text{HE} \times e^{\frac{S^2}{R^2} W_3(z_0)}\,,
 \end{equation}
valid to all orders in $1/R$ in the high energy limit! In particular curvature corrections in the high energy limit exponentiate. Subleading terms, on the other hand, are not captured by this simple model. This is the main result of our paper. This makes a direct connection between the results of \cite{Alday:2023mvu} and classical solutions in $AdS$. Furthermore, it also puts strong constraints on subsequent curvature corrections in the AdS Virasoro-Shapiro amplitude. The rest of the paper is organised as follows. In section \ref{sec:Solution} we set up the classical scattering problem in $AdS$ and solve the equations of motion and Virasoro constraints in a $1/R$ expansion. In section \ref{sec:Action} we evaluate the appropriate action on this classical solution, and show that this agrees with the results of \cite{Alday:2023mvu} in a saddle point approximation.  We then conclude with some discussion and a list of open problems. In the appendices we discuss properties of SVMPLs, the main ingredient in the construction of the classical solutions in AdS; the evaluation of leading and sub-leading terms by saddle point approximation; and the computation of some simple prefactors for the bosonic model under consideration.

\section{String scattering on AdS}

\label{sec:Solution}
We consider a classical string scattering problem on $AdS_d$. The relevant Lagrangian is 
\begin{equation}
{\cal L} = \frac{1}{2\pi} \partial X^M \bar \partial X_M+ \Lambda(X^M X_M +R^2) - i \sum_{k=1}^4 P_k^M  X_M \delta^{(2)}(\zeta-z_k)\,,
\end{equation}
where we parametrise $AdS_{d}$ in terms of embedding coordinates $X^M$, with $M=0,\cdots,d$. We denote the world-sheet coordinates by $(\zeta,\bar \zeta)$ and the location of the punctures by $z_k$, which we assume to be real. The conformal gauge equations of motion arising from the Lagrangian are supplemented by the Virasoro constraints
\begin{equation}
\partial X^N   \partial X_N=\bar \partial X^N  \bar \partial X_N=0\,.
\end{equation}
Away from the punctures $\zeta=z_k$ the equations of motion are
\begin{equation}
\partial \bar \partial X^M = \frac{\partial X^N  \bar \partial X_N}{R^2} X^M,
\end{equation}
with the following boundary conditions as we approach each of the punctures
\begin{equation}
X^M =- i P^M_k \log \left|1-\frac{\zeta}{z_k}\right| + Q^M_k + \cdots\,.
\label{bc}
\end{equation}
The quadratic constraint $X^M X_M =-R^2$ then implies that for each puncture
\begin{equation}
\label{puncturerelations}
P_k^M P_{k,M} = P_k^M Q_{k,M} =0\,.
\end{equation}
The scattering problem in flat space considered in \cite{Gross:1987kza,Gross:1987ar,Gross:1988ue} arises as a limit of the $AdS$ problem as follows. We split the indices into $M=(0,\mu)$ with  $\mu=1,\cdots,d$ and such that 
\begin{equation}
X^M X_M = -X^{0}X^{0} + X^\mu X_\mu \equiv  -X^{0}X^{0}+X \cdot X\,,
\end{equation}
where we reserve the notation $X \cdot X=X^\mu X_\mu$ for the $\mu$-indices. We then consider solutions with a large $R$ expansion 
\bea
X^0 &= R + \frac{1}{R} X^0_1 + \cdots\,,\\
X^\mu &= X^\mu_0+ \frac{1}{R^2} X^\mu_1 + \cdots\,,
\eea{X_exp}
where $X^\mu_0$ is the flat space solution. The corresponding momentum at each puncture $k$ also admits a $1/R$ expansion
\bea
P_k^0 &= \frac{1}{R} p_{k,1}^0 + \cdots\,,\\
P_k^\mu &= p^\mu_{k,0}+ \frac{1}{R^2} p^\mu_{k,1} + \cdots\,.
\eea{P_exp}
As a result of the relations (\ref{puncturerelations}) $p^\mu_{k,0}$ is null from the $d-$dimensional flat space point of view, which is precisely the condition for massless scattering in flat space. Furthermore, solving for $X^0 = \sqrt{R^2+X \cdot X}$, the equations of motion away from the punctures and Virasoro constraints for the flat space coordinates $X_0^\mu$ are simply

\begin{equation}
\partial \bar \partial X_0^\mu = 0,~~~~~\partial X_0 \cdot \partial X_{0}=\bar \partial X_0 \cdot \bar \partial X_{0}=0\,.
\end{equation}
Together with the boundary conditions, the equations of motion imply
\begin{equation}
X^\mu_0 =- i \sum_{k} p_{k,0}^\mu \log \left|1- \frac{\zeta}{z_k}\right| + q_{0}^\mu\,,
\end{equation}
for any constant $q_{0}^\mu$. Plugging this into the Virasoro constraints at leading order
we learn the locations of the punctures are related to the Mandelstam variables of the scattering process
\begin{equation}
\frac{(z_1-z_3)(z_2-z_4)}{(z_1-z_2)(z_3-z_4)}=-\frac{T}{S}\,,
\end{equation}
where $p_{1,0} \cdot p_{2,0}=-2S$, $p_{1,0} \cdot p_{3,0}=-2T$, $p_{1,0} \cdot p_{4,0}=-2U$ with $S+T+U=0$. This of course agrees with the solution found in flat space \cite{Gross:1987kza}. 

\subsection{Higher orders}
We would like to solve the equations of motion and Virasoro constraints in a $1/R$ expansion
\beq
X^\mu = X^\mu_0+ \frac{1}{R^2} X^\mu_1 + \cdots\,.
\eeq
To understand the systematics of this expansion, note that the flat space solution $X^\mu_0$ is single-valued as we move around each puncture $\zeta=z_k$ in the $\zeta-$plane. Indeed, the solution can be written in terms of single-valued multiple polylogarithms (SVMPLs), whose letters are the locations of the punctures. More specifically
\begin{equation}
X^\mu_0 =- \frac{i}{2} \sum_{k} p_{k,0}^\mu {\cal L}_{z_k}(\zeta)\,.
\end{equation}
Consider now the equations of motion at the next order
\begin{equation}
\partial \bar \partial X^\mu_1 = \partial X_0 \cdot  \bar \partial X_0 \, X_0^\mu =  \frac{i}{8} \sum_{i,j,k} \frac{p_{i,0} \cdot p_{j,0}}{(\zeta-z_i)(\bar \zeta-z_j)} p_{k,0}^\mu {\cal L}_{z_k}(\zeta)\,.
\end{equation}
There is a systematic procedure to `integrate' the r.h.s in terms of SVMPLs. For the holomorphic derivative this is very easily done  
\begin{equation}
\int d \zeta \frac{{\cal L}_{w}(\zeta)}{(\zeta-z_i)} \to {\cal L}_{z_i w}(\zeta)\,.
\end{equation}
For the anti-holomorphic derivative it is slightly more complicated, but a recursive algorithm has been worked out in \cite{Brown2004a,Brown2004b,Vanhove:2018elu}, and it is explained in appendix \ref{App:SVMPLs}. The result always takes the form
\begin{equation}
\int d\bar \zeta \frac{{\cal L}_{w}(\zeta)}{(\bar \zeta-z_j)} \to {\cal L}_{w z_j}(\zeta) + \cdots\,,
\end{equation}
where $\cdots$ is given by a sum of terms of uniform weight $|w|+1$. For instance, for $|w|=1$ we obtain
\begin{equation}
\int d\bar \zeta \frac{{\cal L}_{z_k}(\zeta)}{(\bar \zeta-z_j)} \to {\cal L}_{z_k z_j}( \zeta) +    {\cal L}_{z_k}(z_j) {\cal L}_{z_j}( \zeta)- {\cal L}_{z_j}(z_k) {\cal L}_{z_k}( \zeta)\,.
\end{equation}
This allows us to write 
\begin{equation}
 X^\mu_1 = \frac{i}{8} \sum_{i,j,k=1}^4 p_{i,0} \cdot p_{j,0} \  p_{k,0}^\mu \left(
{\cal L}_{z_i z_k z_j}( \zeta) +    {\cal L}_{z_k}(z_j) {\cal L}_{z_i z_j}( \zeta)- {\cal L}_{z_j}(z_k) {\cal L}_{z_i z_k}( \zeta)
\right)\,,
\end{equation}
so that the equations of motion are satisfied. Note that at each order we have the freedom to add weight one functions $ {\cal L}_{z_i}( \zeta)$ in $\zeta$. At each step we use this freedom to remove all weight one functions from the solution. This procedure can be repeated to higher orders and the solution has the following schematic form
\begin{equation}
X^\mu = {\cal L}_1( \zeta) + \frac{1}{R^2}  {\cal L}_3( \zeta)+ \frac{1}{R^4}  {\cal L}_5( \zeta)+ \cdots\,,
\end{equation}
where ${\cal L}_n( \zeta)$ are linear combinations of pure SVMPLs of weight $n$, with either $\zeta$ or $z_i$ as their arguments\footnote{More precisely, by pure we mean that the entire $\zeta$-dependence is through the SVMPLs, and not, for instance, through rational functions multiplying those.}, and letters from the alphabet $\{z_1,z_2,z_3,z_4\}$. The relation $X^0 = \sqrt{R^2+X \cdot X}$ implies a very similar structure for $X^0$:
\begin{equation}
X^0 =R {\cal L}_0(\zeta) + \frac{1}{R}  {\cal L}_2(\zeta) + \frac{1}{R^3}  {\cal L}_4(\zeta) + \cdots\,,
\end{equation}
with ${\cal L}_0(\zeta)=1$. A salient feature of our method, is that once the `seed' solution $X^\mu_0$ is given, the whole tower in $1/R$ is fixed by the equations of motion and the integration procedure described above. The solution has two additional features. First, note that SVMPLs of weight higher than zero are defined so that they vanish at the base point $\zeta=0$. This in particular means that the solution we just constructed satisfies
\begin{equation}
X^0(0)=R,~~~X^\mu(0)=0\,.
\end{equation}
The second condition is the statement that $\partial X^M(0)$ and $\bar \partial X^M(0)$ do not receive $1/R$ corrections. For $\partial X^M(0)$ this is easy to see, since at each order weight one SVMPLs with argument $\zeta$ are removed, consistently with the equations of motion. For $\bar \partial X^M(0)$ this is also true, but one needs to use the fact that $z_i$ are real (which was an assumption in constructing our solutions). In the next section we will see that with these conditions it can be shown that our solution satisfies the Virasoro constraints. 

\subsection{Virasoro constraints}
\label{sec:virasoro}

The structure of the solutions in a $1/R$ expansion implies the following
\beq
\partial X^M = \sum_{k=1}^4 \frac{H_k^M}{\zeta-z_k}\,,~~~\bar \partial X^M = \sum_{k=1}^4 \frac{\bar H_k^M}{\bar \zeta-z_k}\,,
\eeq
where $H_k^M$, $\bar H_k^M$ for $k=1,2,3,4$ are pure SVMPLs in $\zeta$, of higher and higher weight in a $1/R$ expansion. Let us study these objects in more detail. First, they satisfy
\begin{equation}
\label{Vconservation}
\sum_{k=1}^4 H_k^M= \sum_{k=1}^4 \bar H_k^M =0\,,
\end{equation}
which is simply the statement that $\partial X^M$ and $\bar \partial X^M $ do not have a pole at infinity. Furthermore $X^M X_M  = -R^2$ implies $X_M \partial X^M =X_M \bar \partial X^M  =0$, so that 
\begin{equation}
X_M H^M_k =X_M \bar H^M_k =0\,.
\end{equation}
Let us now write the equations of motion in terms of the quantities $H_k^M$, $\bar H_k^M$. There are two equivalent ways of writing them, depending on the order in which we take the derivatives. For instance, one can derive 
\begin{equation}
\sum_k \frac{\bar \partial H_k^M}{\zeta-z_k} = \frac{1}{R^2} \sum_{i,j} \frac{H_i^N \bar H_{j \, N}}{(\zeta-z_i)(\bar \zeta-z_j)} X^M\,.
\end{equation}
Focusing on a given pole $\zeta=z_k$ we obtain
\begin{equation}
\bar \partial H_k^M = \frac{1}{R^2} \sum_{j} \frac{H_k^N \bar H_{j \, N}}{(\bar \zeta-z_j)} X^M\,.
\end{equation}
We can now consider the contractions 
\begin{equation}
S_{k,k'}= H_k^M H_{k' \, M}\,.
\end{equation}
As a consequence of the equations of motions we find that $S_{k,k'}$ is actually holomorphic $\bar \partial S_{k,k'}=0$. On the other hand, our procedure implies $S_{k,k'}$ is given by SVMPLs in $\zeta$. The only SVMPL which is also holomorphic is the constant function, so that
\begin{equation}
S_{k,k'} \text{ is independent of $\zeta,\bar \zeta$}\,.
\end{equation}
This result is valid to all orders in $1/R$. For the solution computed above, we can actually do better and compute these products exactly. In our procedure $H_k^M$, $\bar H_k^M$ are given in terms of SVMPLs, where the point $\zeta=0$ has been chosen as the base point. In particular this implies that we can compute $H_k^M$, $\bar H_k^M$ exactly at that point. Indeed
\begin{equation}
H_k^\mu(0) = \bar H_k^\mu(0) = -\frac{i}{2} p_{k,0}^\mu\,,~~~H_k^0(0) = \bar H_k^0(0) =0\,.
\end{equation}
Since $S_{k,k'}$ is constant, we can evaluate it at any point, and in particular at $\zeta=0$, where we obtain $S_{k,k'}=-\frac{p_{k,0} \cdot p_{k',0}}{4}$. Note in particular $S_{k,k}=0$. Furthermore we also have $S_{1,2}+S_{1,3}+S_{1,4}=0$, which follows from (\ref{Vconservation}) together with $S_{k,k}=0$. The same considerations follow for the contractions  $\bar S_{k,k'}= \bar H_k^M \bar H_{k' \, M}$. With this structure, the Virasoro constraints take exactly the same form as in flat space! In particular we obtain a single constraint on the location of the punctures which is now
\begin{equation}
\frac{(z_1-z_3)(z_2-z_4)}{(z_1-z_2)(z_3-z_4)}=-\frac{p_{1,0} \cdot p_{3,0}}{p_{1,0} \cdot p_{2,0}}\,,
\end{equation}
exactly as in flat space. This proves that our construction in terms of SVMPLs satisfies both, the equations of motion as well as the Virasoro constraints. In the next section, we consider deformations of our solution, consistent with both, the equations of motion and the Virasoro constraints.   

\subsection{Higher order momenta and deformations}
Given a seed solution $X_0^\mu$, the procedure described above gives a solution in a $1/R$ expansion, to arbitrary order
\begin{equation}
X_0^\mu \to X^\mu = X_0^\mu + \frac{1}{R^2}X_1^\mu+\cdots\,.
\end{equation}
Looking at the behaviour around each puncture, this determines the momenta in a $1/R$ expansion from the seed/flat space momenta $p_{k,0}^\mu$:
\begin{equation}
p_{k,0}^\mu \to P_k^\mu = p^\mu_{k,0}+ \frac{1}{R^2} p^\mu_{k,1} + \cdots.
\end{equation}
In the following we will consider a family of deformations of the solution, consistent with the equations of motion and Virasoro constraints, such that the flat space momenta $p_{k,0}^\mu$ are invariant. The first deformation is labelled by a constant vector $W^\mu$ and reduces to translations in the flat space limit. It acts on coordinates as $X^\mu \to \hat X^\mu$ and $X^0 \to \hat X^0$ with 
\begin{align}
\hat X^\mu &= X^\mu + W^\mu \frac{X^0}{R}+ W^\mu W \cdot X \frac{\sqrt{1+W^2/R^2}-1}{W^2}\,, \label{XmuHat}\\
\hat X^0&=\sqrt{1+W^2/R^2}\left(X^0 + \frac{ W \cdot X}{\sqrt{R^2+W^2}}\right)\,.
\end{align}
In other words, it is a Lorentz transformation on the embedding coordinates
\begin{equation}
X^M \to \hat X^M = \Lambda^M_N X^N\,,
\end{equation}
with $ \Lambda^M_N  \eta_{MM'}  \Lambda^{M'}_{N'} = \eta_{NN'}$. As such it preserves the inner products $S_{k,k'}$, as well as the equations of motion. As already mentioned, in the flat space limit this transformation reduces to shifts by $w^\mu=W^\mu(R=\infty)$, and hence the flat space momenta are invariant. On the other hand, it acts on the momenta at higher orders in $1/R$. We will fix the freedom implied by this deformation as follows. While the flat space/seed momenta are conserved $\sum_k p_{k,0}^\mu=0$, this is not true at higher orders in $1/R$. Since we are considering a scattering problem around flat space, where the coordinates $\mu-$parametrise this space, it seems reasonable to impose momentum conservation on the $\mu-$plane. We can use the $W-$deformation to achieve that, so that the total momentum in the $\mu-$ coordinates is conserved $\sum_k \hat P^\mu_k=0$. Let us discuss this in detail. Imagine we carry out the procedure described previously. This will lead to   $P_{T}^\mu = \sum_k P^\mu_k$ which is non-zero from order $1/R^2$. Perform now a transformation such that
\begin{equation}
\hat P_{T}^\mu = P_{T}^\mu + W^\mu \frac{P_{T}^0}{R}+ W^\mu W \cdot P_{T} \frac{\sqrt{1+W^2/R^2}-1}{W^2} =0 
\end{equation}
This can be achieved by choosing  (for $P_{T}^0>0,(P_{T}^0)^2-P_{T}^\mu P_{T}^\mu>0$)
\begin{equation}
W^\mu = -\frac{R P_{T}^\mu }{\sqrt{(P_{T}^0)^2-P_{T}^\mu P_{T}^\mu}}\,.
\label{W_choice}
\end{equation}
To leading order this gives
\beq
W^\mu = \frac{i}{16} \sum\limits_{\substack{i,j,k=1\\i\neq j, i\neq k, j \neq k}}^4
p_{i,0}^\mu
\left(
\mathcal{L}_{z_i}(z_j) - \mathcal{L}_{z_j}(z_k) 
\right) + O\left(1/R^2 \right)\,.
\eeq
On the other hand, note that the momentum will not be conserved in the $0$-direction. Indeed, this cannot be achieved, as translations are not a symmetry of $AdS$. Note also the following. While the embedding momentum at each puncture is null $P_k^M P_{k \,M}=0$, from order $1/R^2$ there is an `induced' mass $(P^0_k)^2= P_k \cdot P_{k}$. It turns out that for the choice in which the momentum along the $\mu-$coordinates is conserved, this induced mass does not depend on the puncture $P_k \cdot P_{k} = -m^2$.
In the remainder of the paper we will use the transformed solution \eqref{XmuHat} with $W^\mu$ given in \eqref{W_choice}.

Let us discuss another type of deformation where we rescale the seed momenta by a factor $p_0^\mu \to \lambda p_0^\mu$. This acts in a very simple way at higher orders in $1/R$. More specifically, at each puncture 
\begin{equation}
p^\mu_0 + \frac{1}{R^2} p^\mu_1+ \frac{1}{R^4} p^\mu_2 + \cdots \to \lambda p^\mu_0 + \frac{\lambda^3}{R^2} p^\mu_1+ \frac{\lambda^5}{R^4} p^\mu_2 + \cdots\,.
\label{rescale_momenta}
\end{equation}
At the level of the Mandelstam invariants this rescales $S \to \lambda^2 S$. In particular, if $\lambda=1+\frac{\alpha}{R^2} + \cdots$ it modifies the Mandelstam invariants as $S \to (1+ 2 \frac{\alpha}{R^2} + \cdots) S$. Note that ratios of Mandelstam invariants are invariant. On the solutions $X^\mu(R,p_0)$, where $p_0$ denotes the seed flat-space momenta, this acts as
\begin{equation}
X^\mu(R,\lambda p_0) = \lambda X^\mu(R/\lambda, p_0)
\quad \Leftrightarrow \quad
X^\mu(\lambda R, \lambda p_0) = \lambda X^\mu( R, p_0)\,,
\end{equation}
as expected.

\section{Evaluating the action}
\label{sec:Action}
Having found the solution in a $1/R$ expansion, we need to evaluate the action at this classical solution. In order to make contact with flat space, while still having a finite answer, we need to subtract a constant and the Lagrangian to consider is
\begin{equation}
{\cal L} = \frac{1}{2\pi} \partial X^M \bar \partial X_M-i \sum_{k=1}^4 \left( P_k^M  X_M +P_k^0 R \right)\delta^{(2)}(\zeta-z_k) =  \frac{1}{2\pi} \partial X^M \bar \partial X_M-i R \sum_{k=1}^4 P_k^0\delta^{(2)}(\zeta-z_k)
\end{equation}
where note that in our signature $X_0=-X^0=-R + \cdots$. In the second expression we took into account that $P_k^M  X_M \to 0$ as we approach the puncture, so its contribution vanishes. In other words, the total contribution to the action is given by the bulk action plus the sum over the `masses' $R m= i R P_k^0$, which, as discussed in the previous section, is independent of the puncture. The contribution from both terms is finite. Indeed, $P_k^0 R$ is clearly finite while the bulk contribution is integrable. Indeed, the non-integrable divergence $\partial X^M \bar \partial X_M \sim \frac{P_k^M P_{k \, M}}{|\zeta-z_k|^2}$ as we approach the puncture vanishes, since $P_k^M P_{k \, M}=0$. To recover flat space we solve for $X^0 = \sqrt{R^2+X \cdot X}$ and expand in $1/R$
\begin{equation}
{\cal L} = \frac{\partial X \cdot 
\bar \partial X}{2\pi}  - i \hspace{-1pt}  \sum_{k=1}^4 P_k \cdot X \delta^{(2)}(\zeta-z_k)- \frac{1}{R^2} \hspace{-2pt} \left( 
 \frac{X \cdot \partial X X \cdot \bar \partial X}{2\pi}  - \frac{i}{2} \hspace{-1pt} \sum_{k=1}^4 p_{k,1}^{0} X \cdot X  \delta^{(2)}(\zeta-z_k) \right) + \ldots
\end{equation}
At leading order we obtain the usual Lagrangian in flat space. 
We will separately evaluate the bulk and vertex operator terms of the action on the classical solution
\bea
{\cal S} ={}& {\cal S}_\text{bulk} + {\cal S}_\text{vertex}\,,
\eea{action}
where
\begin{align}
{\cal S}_\text{bulk} ={}& \frac{1}{2\pi} \int d^2 \zeta \, \left( \partial X \cdot \bar \partial X - \frac{1}{R^2} X \cdot \partial X X \cdot \bar \partial X    + \ldots\, \right), \label{action_expanded} \\
{\cal S}_\text{vertex} ={}&  
-i \sum_{k=1}^4 \left(p_{k,1}^0 + \frac{1}{R^2} p_{k,2}^0  + \ldots\right).
\nonumber
\end{align}
We will denote different terms in the expansion by
${\cal S} = {\cal S}^{(0)} + \frac{1}{R^2}  {\cal S}^{(1)} + \cdots$.
The first two terms in the bulk action are given by
\begin{align}
{\cal S}^{(0)}_\text{bulk} ={}& \frac{1}{2\pi} \int d^2 \zeta \,  \partial X_0 \cdot \bar \partial X_0 = \int d^2 \zeta \, \sum_{i,j=1}^{4} \frac{c_{ij}}{(\zeta-z_i)(\bar{\zeta}-z_j)}\,,\label{Sbulk_terms}\\
{\cal S}^{(1)}_\text{bulk} ={}& \frac{1}{2\pi} \int \hspace{-3pt} d^2 \zeta \left( \partial X_0 \cdot \bar \partial X_1 + \partial X_1 \cdot \bar \partial X_0 - X_0 \cdot \partial X_0 X_0 \cdot \bar \partial X_0 \right)
=\int \hspace{-3pt} d^2 \zeta \hspace{-2pt} \sum_{i,j=1}^{4} \frac{f_{ij}(\zeta)}{(\zeta-z_i)(\bar{\zeta}-z_j)}\,, \nonumber
\end{align}
where $c_{ij}$ are rational numbers and $f_{ij}(\zeta)$ is of transcendental weight two and can be written in terms of SVMPLs.
The integral can be done using the formula \eqref{sphere_integration}, where the residues at $\bar{\zeta}=\infty$ cancel in the sums. The results, expressed in terms of the cross-ratio $z_0=-\frac{z_{12} z_{34}}{z_{23}z_{14}}$, read
\begin{align}
{\cal S}^{(0)}_\text{bulk} ={}& -S \left(\mathcal{L}_0(z_0) + \frac{1-z_0}{z_0} \mathcal{L}_1(z_0)\right)\,, \label{Sbulk_results}\\
{\cal S}^{(1)}_\text{bulk} ={}& S^2 \bigg(
\frac{2-z_0}{z_0}\mathcal{L}_{001}\left(z_0\right)
+\frac{z_0-2 }{z_0}\mathcal{L}_{010}\left(z_0\right)
+\frac{1-z_0^2}{z_0^2} \mathcal{L}_{011}\left(z_0\right)
+\frac{z_0^2-1}{z_0^2} \mathcal{L}_{101}\left(z_0\right)
+6 \zeta (3)
\bigg)\,. \nonumber
\end{align}

In order to evaluate the source terms, we can simply use the relation $P_k^M Q_{k,M} = 0$ to write $P^0_k$ in terms of $P^\mu_k$ and $Q^\mu_k$, which can be read off from $X^\mu$ using \eqref{bc} and \eqref{L_sing}.
We have
\beq
p^0_{k,1} = p_{k,0} \cdot q_{k,0}\,, \qquad
p^0_{k,2} = p_{k,0} \cdot q_{k,1}+ p_{k,1} \cdot q_{k,0} - \frac12 q_{k,0} \cdot q_{k,0} p_{k,0} \cdot q_{k,0} \,,
\eeq
leading to the expressions
\begin{align}
{\cal S}^{(0)}_\text{vertex} ={}& 2S \left(\mathcal{L}_0(z_0) + \frac{1-z_0}{z_0} \mathcal{L}_1(z_0)\right) \,, \nonumber\\
{\cal S}^{(1)}_\text{vertex} ={}& S^2 \bigg(
\frac{1}{2} \mathcal{L}_{000}\left(z_0\right)
+\frac{\left(z_0-3\right)  }{z_0}\mathcal{L}_{001}\left(z_0\right)
-\frac{\left(3 z_0-5\right) }{2 z_0}\mathcal{L}_{010}\left(z_0\right)\label{Ssource_results}\\
&+\frac{\left(z_0^2+z_0-2\right) }{z_0^2}\mathcal{L}_{011}\left(z_0\right)
-\frac{\left(3 z_0^2-z_0-2\right)  }{2 z_0^2}\mathcal{L}_{101}\left(z_0\right)
+\frac{\left(z_0-1\right){}^2 }{2 z_0^2}\mathcal{L}_{111}\left(z_0\right)
-8 \zeta (3)
\bigg) \,.\nonumber
\end{align}
In summary, this classical solution predicts the following behaviour of the amplitude in the high-energy limit
 \begin{equation}
 A_{4,\text{bos}}^{AdS}(S,T)_\text{HE} \sim e^{-\mathcal{S}} =  e^{S V_1(z_0) + \frac{S^2}{R^2} V_3(z_0) +\frac{S^3}{R^4} V_5(z_0) +\cdots}\,.
\label{A_bos}
 \end{equation}
In addition, as discussed above around \eqref{rescale_momenta}, we have still the freedom to rescale the momenta such that $S \to S(1 + \frac{S F_2(z_0)}{R^2} + \cdots)$, while keeping ratios $S/T$ invariant. Although these corrections are suppressed in the regime we are studying (large $S,R$ with fixed $S^2/R^2$), the first correction does enter the leading high-energy limit of the amplitude via the term $S V_1(z_0)$ in the exponential. After performing this rescaling, we have
\beq
S V_1 (z_0) = -{\cal S}^{(0)}\,, \qquad
S^2 V_3 (z_0) = -{\cal S}^{(1)} - 2 S F_2(z_0) {\cal S}^{(0)}\,,
\eeq
while $V_5 (z_0)$ and further terms do not contribute to the leading high-energy limit. While $F_2(z_0)$ cannot be determined purely within the context of this classical bosonic model (being a subleading quantity), it will be fixed by the comparison below. 

Now we would like to compare this to the high-energy limit of the AdS Virasoro-Shapiro amplitude computed in \cite{Alday:2023jdk,Alday:2023mvu}. The AdS amplitude is given in terms of a  world-sheet integral representation, identical to that for flat space, with the extra insertion of single-valued functions $W_n(z)$. Order by order in $1/R$ these extra insertions depend on $S,T$ polynomially, so that the location of the saddle point is the same as in flat space. The high energy limit, in a $1/R$ expansion, is then obtained by simply evaluating the world-sheet integral representation on the saddle point $z = z_0$ and takes the form
 \begin{equation}
 \label{hedirect2}
 A_{4}^{AdS}(S,T)_\text{HE} \sim  e^{-2 S \log|S| -2 T \log|T| - 2U \log |U|}  \left(1 + \frac{S^2}{R^2}W_3(z_0) + \frac{S^4}{R^4}W_6(z_0) +\cdots \right)\,,
 \end{equation}
where
\bea
W_3 (z) &= \frac{U^2}{S^2} G^{(1)}_\text{tot}(S,T,z)\,,\\
W_6 (z) &= \frac{U^2}{S^4} G^{(2)}_\text{tot}(S,T,z)\,, 
\eea{W36_def}
with $G^{(1,2)}_\text{tot}(S,T,z)$ as in \cite{Alday:2023mvu}.
On the saddle point we find
\bea
W_3 (z_0) ={}&
\mathcal{L}_{000}\left(z_0\right)
-\mathcal{L}_{001}  \left(z_0\right)
-\frac{ 1}{z_0}\mathcal{L}_{010}\left(z_0\right)
-\frac{\left(z_0-1\right){}^2}{z_0^2}\mathcal{L}_{011}\left(z_0\right)\\
&+\frac{\left(z_0-1\right)    }{z_0^2}\mathcal{L}_{101}\left(z_0\right)
+\frac{\left(z_0-1\right){}^2 }{z_0^2}\mathcal{L}_{111}\left(z_0\right)
+2 \zeta (3)\,,
\eea{W3}
and
\beq
W_6 (z_0) = \frac12 W_3 (z_0)^2\,.
\label{W6_saddle}
\eeq
We are now ready to compare \eqref{A_bos} with \eqref{hedirect2}.
The leading term of course reproduces the flat space result
\beq
e^{S V_1(z_0)}
= e^{-2 S \log |S|-2 T \log |T|-2 U \log |U|}\,.
\eeq
More interestingly, the full high-energy limit of the AdS Virasoro-Shapiro amplitude to all orders in $\frac{S^2}{R^2}$ is determined by the subleading exponent, and it precisely matches the expected result
\beq
\label{exp}
e^{\frac{S^2}{R^2} V_3(z_0)} = \left(1 + \frac{S^2}{R^2}W_3(z_0) + \frac{S^4}{R^4}W_6(z_0) +\cdots \right)\,,
\eeq
provided we choose
\beq
F_2(z_0) = \frac{1}{4} \left(-  \mathcal{L}_{00}\left(z_0\right)
+\frac{2}{z_0} \mathcal{L}_{01}\left(z_0\right)
+\frac{ z_0-1}{z_0} \mathcal{L}_{11}\left(z_0\right) \right)\,.
\eeq
Note that it is quite non-trivial that such an $F_2(z_0)$ exists at all. This happens because the combination $S^2 W_3(z_0)+{\cal S}^{(1)}$ neatly factorises into something of weight two times ${\cal S}^{(0)}$. Furthermore the bosonic model predicts exponentiation to all orders in $S^2/R^2$, via (\ref{exp}). The highly non-trivial relation \eqref{W6_saddle} is a test of exponentiation to quadratic order.

\subsection{Universality}
\label{sec:universality}
The AdS Virasoro-Shapiro amplitude in the high energy limit, namely large $R,S,T$ with fixed $S/R$ and $S/T$, takes the form 
\begin{equation}
A_{4}^{AdS}(S,T)_\text{HE}  = A_{4}^\text{flat}(S,T)_\text{HE} \times e^{\frac{S^2}{R^2} W(S/T)} \,.
\end{equation}
This exponentiation can be understood in terms of a bosonic model describing the scattering of classical strings on $AdS$
 \begin{equation}
 A_{4,\text{bos}}^{AdS}(S,T)_\text{HE} \sim e^{-\mathcal{S}}\,,
 \end{equation}
evaluated on the classical solution $-\mathcal{S}=S V_1(z_0) + \frac{S^2}{R^2} V_3(z_0) +\frac{S^3}{R^4} V_5(z_0) +\cdots$. The universality of our results can be understood as follows. Note that at the level of the action the contribution we are computing is much smaller than the flat space contribution, indeed  $S V_1(z_0) \gg \frac{S^2}{R^2} V_3(z_0)$, but we keep it because it is in the exponential. Higher order curvature corrections, on the other hand, can be safely ignored. In other words, in the regime we are considering, only the first order curvature corrections around flat space are important.\footnote{We thank J. Maldacena for discussions on this point.}

Related to the universality of our results, let us discuss two ingredients that were not taken into account in the bosonic model we considered. The first is the specific form of the $AdS$ vertex operators. In our computation we assumed that the overall prefactor is that of flat space. As we take into account curvature corrections this will get modified and on dimensional grounds we expect the following schematic form (see appendix C for more details)
 \begin{equation}
\text{pref}_{AdS} = \text{pref}_\text{flat} \left(1+ \frac{(X^{(0)}_\text{classical})^2}{R^2} + \cdots \right)\,.
 \end{equation}
But since $(X^{(0)}_\text{classical})^2 \sim S$, this and higher order corrections can be disregarded in the regime under consideration. Finally, we have also ignored quantum corrections, including contributions from fermionic fields. These corrections can indeed affect our results and are encapsulated in the `non-universal' function $F_2(z_0)$, given above. It would be very interesting to derive $F_2(z_0)$ directly from a world-sheet computation. At present this is unavailable.

\section{Conclusions}

In this paper we have computed the AdS Virasoro Shapiro amplitude, for the scattering of four gravitons in $AdS_5 \times S^5$ at tree level, in the fixed-angle high energy limit. This limit provides a regime where the amplitude can be computed to all orders in the curvature expansion with the result
\begin{equation}
A_{4}^{AdS}(S,T)_\text{HE}  = A_{4}^\text{flat}(S,T)_\text{HE} \times e^{\frac{S^2}{R^2} W_3(z_0)} \,,
\end{equation}
with $W_3(z_0)$ a transcendental function of weight three. This result can be explicitly checked to order $1/R^4$ against the results of \cite{Alday:2023mvu} and is reproduced to all orders by a classical scattering computation in $AdS$. This result opens the path to a more direct comparison to a putative world-sheet description of strings on $AdS_5 \times S^5$. Perhaps the first step to explore this is to use the RNS formalism to compute the AdS amplitude in a curvature expansion. It would also be interesting to understand whether/how the pure spinor formalism simplifies at high energies, and whether it can be used to compute subleading corrections. There are many other directions that would be worth exploring. 

The results of this paper can be fed into the program of  \cite{Alday:2023mvu}, and put strong constraints on curvature corrections at higher orders, in particular fixing the answer at the saddle point. It would be interesting to see if this can be combined with results from integrability and the structure of the solution to go to higher orders. It would also be interesting to see if sub-leading corrections to the high energy limit also follow a simple pattern. It is straightforward to extend the computation of the bosonic model to higher point functions. The location of the punctures serve as the letters of the SVMPLs from which the solution will be constructed. Furthermore we expect again that in a $1/R$ expansion the location of the saddle point is the same as for flat space. It would also be interesting to study the bosonic model exactly, for finite $R$, perhaps by using a Pohlmeyer-type reduction \cite{Pohlmeyer:1975nb}, as was done for the problem of scattering amplitudes in MSYM at strong coupling \cite{Alday:2009yn}.  Finally, a fascinating feature of the world-sheet computation is that one can go to higher genus, and the genus expansion is Borel summable \cite{Mende:1989wt}. It would be very interesting to explore this question in $AdS$.

\section*{Acknowledgements} 

We thank Xinan Zhou for collaboration at early stages of this project.
The work of LFA and TH is supported by the European Research Council (ERC) under the European Union's Horizon
2020 research and innovation programme (grant agreement No 787185). LFA is also supported in part by the STFC grant ST/T000864/1.
TH is also supported by the STFC grant ST/X000591/1.

\appendix

\section{Single-valued multiple polylogarithms}

\label{App:SVMPLs}

The classical solutions considered in this paper are built up from single-valued multiple polylogarithms. These are built from multiple polylogarithms (MPLs) which we introduce first. They are functions of one variable $L_w(z)$ and for the purposes of this paper they are labelled by words in the alphabet $\{ \sigma_1, \sigma_2,\sigma_3, \sigma_4 \}$. They are recursively defined as follows
\begin{equation}
\label{recursionL}
\partial L_{\sigma_i w}(z) = \frac{L_w(z)}{z-z_i},
\end{equation}
together with $\lim_{z \to 0} L_w(z)=0$ except for  $L_e(z)=1$ where $e$ denotes the empty word. Here $z_i$ denote the location of the punctures, which are taken to be real and non-zero. For instance
\begin{equation}
\partial L_{\underbrace{\sigma_i \cdots \sigma_i}_{p \text{ times}}}(z) =\frac{\log^p\left(1-\frac{z}{z_i} \right)}{p!}\,.
\end{equation}
It is convenient to introduce the generating function
\begin{equation}
L_\sigma(z) = 1+ \sum_{i=1}^4 L_{\sigma_i}(z) \sigma_i+ \sum_{i,j=1}^4 L_{\sigma_i \sigma_j}(z) \sigma_i \sigma_j + \cdots\,.
\end{equation}
In terms of this the recursion relation (\ref{recursionL}) can be written as
\begin{equation}
\partial L_\sigma(z) = \sum_{i=1}^4 \frac{\sigma_i}{z-z_i}L_\sigma(z) \,.
\end{equation}
MPLs are not single-valued as we go around the punctures. In order to construct single-valued multiple polylogarithms (SVMPLs) we follow \cite{Brown2004a,Brown2004b,Vanhove:2018elu}. Consider the map that reverses the order of letters in a word
\begin{equation}
w = \sigma_{i_1} \cdots  \sigma_{i_r} \to \widetilde{w} = \sigma_{i_r} \cdots  \sigma_{i_1}\,.
\end{equation}
Then SVMPLs are defined as the coefficients ${\cal L}_w(z)$ in the following generating series
\begin{equation}
{\cal L}_\sigma(z)=  L_\sigma(z) \widetilde{\left( \bar L_{\sigma'}(z) \right)}\,,
\end{equation}
where the letters $\sigma'_i$ are related to the letters $\sigma_i$ by
\begin{equation}
\sigma_i' = {\cal L}_\sigma(z_i)^{-1} \sigma_i {\cal L}_\sigma(z_i) = \sigma_i + \sum_{j=1}^4 {\cal L}_{\sigma_j}(z_i)\left( \sigma_i \sigma_j - \sigma_j \sigma_i \right) + \cdots\,.
\end{equation}
With this relation worked out, we can then determine the action of holomorphic and anti-holomorphic derivatives. Indeed
\beq
\partial {\cal L}_\sigma(z) = \sum_{i=1}^4 \frac{\sigma_i}{z-z_i}{\cal L}_\sigma(z),~~~\bar \partial {\cal L}_\sigma(z) = {\cal L}_\sigma(z)  \sum_{i=1}^4 \frac{\sigma'_i}{\bar z-z_i}\,.
\eeq
For instance, for weight two the second relation implies
\beq
\bar \partial {\cal L}_{\sigma_i \sigma_j}(z) = \frac{{\cal L}_{\sigma_i}(z)}{\bar z-z_j} + \frac{{\cal L}_{\sigma_j}(z_i)}{\bar z-z_i}- \frac{{\cal L}_{\sigma_i}(z_j)}{\bar z-z_j}\,.
\eeq
SVMPLs satisfy various relations and in particular the shuffle relations for their product
\begin{equation}
{\cal L}_{w}(z) {\cal L}_{w'}(z) = \sum_{W \in w \shuffle w'}{\cal L}_{W}(z) \,.
\end{equation}
These relations can be used to isolate the logarithmic singularities of SVMPLs as the argument approaches the first label. For instance
\beq
\lim\limits_{z \to z_i} \mathcal{L}_{\sigma_i \sigma_j w} (z) = \mathcal{L}_{\sigma_j w} (z_i) \mathcal{L}_{\sigma_i} (z) + \, \text{finite}\,, \qquad i \neq j\,.
\label{L_sing}
\eeq
We found the tools \cite{Duhr:2019tlz,Panzer:2014caa} very useful when working with SVMPLs, in particular for rewriting SVMPLs of $z_1$, $z_2$, $z_3$, $z_4$ in terms of the cross-ratio $z_0$ in section \ref{sec:Action}. Finally in this work we are interested in integrating expressions containing SVMPLs over the Riemann sphere. All expressions are of the form \cite{Vanhove:2018elu}
 \begin{equation}
\frac{1}{2 \pi} \int d^2z \frac{{\cal L}_w(z)}{(z-z_i)(\bar z-z_j)} = - {\cal L}_{\sigma_i w}(z_j) + \underset{\zb = \infty}{\text{Res}}\, \frac{{\cal L}_{\sigma_i w}(z)}{\bar z-z_j}\,.
\label{sphere_integration}
 \end{equation}
The contribution from the point at infinity is slightly involved, however, for the integrals considered in this paper (the bulk part of the action), the total contribution at infinity vanishes.  

\section{Saddle points}
In this paper we compute the fixed-angle high energy limit, ${\it i.e.}$ the large $S,T$ limit with $S/T$ fixed, of expressions of the form
\begin{equation}
F(S,T) = \int d^2 z |z|^{-2S} |1-z|^{-2T} W(z)\,,
\end{equation}
where $W(z)$ depends on $S,T$ in a polynomial way. In the large $S,T$ limit this integral is well approximated by a saddle point. Writing
\begin{equation}
 |z|^{-2S} |1-z|^{-2T} = e^{-2 S \log|z| -2 T \log |1-z|}\,,
\end{equation}
 we see that the saddle point location is at 
 \begin{equation}
 z=\bar z=\frac{S}{S+T} \equiv z_0\,.
 \end{equation}
Expanding around this saddle point $z=z_0+\epsilon$, $\bar z =z_0 + \bar \epsilon$ we obtain
\begin{equation}
F(S,T)_\text{HE} =e^{-2S \log |S| -2 T \log |T|-2U \log |U|} W(z_0) \int d^2\epsilon e^{-\frac{U^3}{2 S T}(\epsilon^2 +\bar \epsilon^2)}\,.
\end{equation}
The integral over $\epsilon,\bar \epsilon$ produces an extra prefactor\footnote{In performing this integral a suitable analytic continuation was assumed.}
\begin{equation}
\int d^2\epsilon e^{-\alpha(\epsilon^2 +\bar \epsilon^2)} = \frac{\pi}{2 \alpha \, i}\,,
\end{equation}
so that 
\begin{equation}
F(S,T)_\text{HE} = \frac{\pi}{i} \frac{S T}{U^3} e^{-2S \log |S| -2 T \log |T|-2U \log |U|} W(z_0)\,.
\end{equation}
We can also compute corrections to this result, with the same exponential behaviour but suppressed by powers of $S,T$. Expanding to higher order in $(\epsilon,\bar \epsilon)$  and keeping only the terms relevant for the first order corrections, we obtain:
\bea
F(S,T)_\text{HE} ={}& \frac{\pi}{2 i \alpha} W(z_0)
+ \frac{\pi}{8 i \alpha^2} \left(\partial^2+\bar \partial^2 \right) W(z_0)
+ \frac{\pi}{8 i \alpha^3} \frac{(S-T) U^4}{S^2 T^2}  \left(\partial+\bar \partial \right) W(z_0) \\
&- \frac{3\pi}{16 i \alpha^3} \frac{U^5 (S^2-S T+T^2)}{S^3 T^3} W(z_0) 
+ \cdots\,,
\eea{saddle_subleading}
where $\alpha=U^3/(2 S T)$ and the holomorphic and anti-holomorphic derivatives are taken before setting $z=\bar z=z_0$. 

\section{Prefactors}
In the body of this paper we have ignored the prefactors in the vertex operators for the bosonic model on $AdS$. The reason is that these prefactors will be evaluated at the classical solution and the difference between the AdS and flat space computations is expected to be of the form 
\begin{equation}
{\cal V}^{AdS}_k= {\cal V}^\text{flat}_k \left( 1+ \frac{c S}{R^2} + \cdots \right)\,,
\end{equation}
and hence they are not relevant at leading order in the high energy limit. Regardless, it is interesting to consider simple prefactors for the bosonic model under consideration. Let us consider for instance the prefactor corresponding to massless excitations 
\begin{equation}
{\cal V}_k = \partial X^M \bar \partial X^N h_{k, M N},~~~\text{at $\zeta=z_k$}\,,
\end{equation}
where the polarisation tensor is perpendicular to the momentum $P_k^M h_{k,M N} = P_k^N h_{k,M N}=0$. $h_{M N}$ has three components. For instance, its symmetric traceless part describes the graviton, while its trace describes the dilaton. Let us consider this component. For each puncture (and suppressing the index $k$) we can take
\begin{equation}
 h_{M N}^{D} = \eta_{MN}- P_M \rho_N - \rho_M P_N\,,
\end{equation}
with $P^M \rho_M=1$ and $\rho^M \rho_M=0$. The first of these conditions ensures the transversality condition, while the second is chosen for convenience. For each puncture we then need to compute
\begin{equation}
{\cal V}_k =\partial X^M \bar \partial X_M - (P_k^M \partial X_M )  (\rho_k^M \bar \partial X_M ) - (\rho_k^M \partial X_M )  (P_k^M \bar \partial X_M )\,,
\end{equation}
at the location of the puncture. To perform this computation it is convenient to write
\begin{equation}
\partial X^M = \sum_{k=1}^4 \frac{H_k^M}{\zeta-z_k}\,,~~~\bar \partial X^M = \sum_{k=1}^4 \frac{\bar H_k^M}{\bar \zeta-z_k}\,.
\end{equation}
We furthermore introduce the following notation for contractions of $H_k$ and $\bar H_k$:
\begin{equation}
\Phi_{k,k'}(\zeta) \equiv H_k^M(\zeta) \bar H_{k' \, M}(\zeta)\,.
\end{equation}
This object is a complicated function, of higher and higher transcendentality in a $1/R$ expansion, but one can explicitly check $\Phi_{k,k}(\zeta)=0$ and $\sum_{k'} \Phi_{k,k'}(\zeta) =\sum_{k} \Phi_{k,k'}(\zeta) = 0$. Furthermore, the matrix with elements $\Phi_{k,k'}(\zeta)$ is Hermitian. After a short computation one can show
\begin{equation}
{\cal V}_k =\sum_{i,j \neq k}^4 \frac{\Phi_{i,j}(z_k)}{(z_k-z_i)(z_k-z_j)}\,.
\end{equation}
This can actually be computed exactly since $\Phi_{k,k'}(\zeta)$ simplifies at the punctures. Let's take $k=1$ for definiteness and note that
\begin{equation}
H_1(z_1) = \bar H_1(z_1) = -\frac{i}{2} P_1\,.
\end{equation}
This then implies
\begin{equation}
\Phi_{1,2}(z_1) = H_1^M(z_1) \bar H_{2 \, M}(z_1) =\bar H_1^M(z_1) \bar H_{2 \, M}(z_1) = S_{1,2}= \frac{S}{2} = \Phi_{2,1}(z_1) \,,
\end{equation}
and in the same way 
\begin{equation}
\Phi_{3,1}(z_1) =\Phi_{1,3}(z_1) =\frac{T}{2},~~~\Phi_{4,1}(z_1)=\Phi_{1,4}(z_1) =\frac{U}{2}\,. 
\end{equation}
Furthermore as $z_1$ is real we have $\Phi_{i,j}(z_1)=\Phi_{j,i}(z_1)$. We have now enough constraints to fix all of them and we find
\begin{equation}
\Phi_{2,3}(z_1) =\Phi_{3,2}(z_1) =\frac{U}{2}\,,~~~\Phi_{2,4}(z_1)=\Phi_{4,2}(z_1) =\frac{T}{2}\,,~~~\Phi_{3,4}(z_1)=\Phi_{4,3}(z_1) =\frac{S}{2}\,.
\end{equation}
These relations are valid to all orders in $1/R$. Similar results can be obtained at all punctures. We can then compute
\begin{equation}
\prod_{k=1}^4 {\cal V}_k = \frac{S^4}{2^{12} (z_1-z_2)^4(z_3-z_4)^4}\,.
\end{equation}
\bibliographystyle{JHEP}
\bibliography{HEDraft}

\end{document}